\documentclass[a4paper,12pt]{article}
\usepackage{a4wide}
\usepackage{times,mathptm}
\usepackage{graphicx}
\newcommand{\beq}{\begin{equation}}
\newcommand{\eeq}{\end{equation}}
\newcommand{\m}{\mu}
\newcommand{\dg}{\dagger}
\newcommand{\ra}{\rightarrow}
\title{\bfseries\Large First Evidence for Center Dominance\\
in SU(3) Lattice Gauge Theory}
\author{\bfseries\large Manfried Faber,$^{a}$ Jeff Greensite,$^{b,c}$
and {\v S}tefan Olejn{\'\i}k$^{d}$\\[2mm]
\normalsize 
$^{a}$Institut f\"ur Kernphysik, Technische Universit\"at Wien,
A--1040 Vienna, Austria;\\[-1mm]
\normalsize 
e-mail: {\tt faber@kph.tuwien.ac.at}\\
\normalsize 
$^{b}$Physics and Astronomy Department, San Francisco
State University,\\[-1mm]
\normalsize 
San Francisco, CA~94117, USA; e-mail: {\tt greensit@quark.sfsu.edu}\\
\normalsize 
$^{c}$Theory Group, Lawrence Berkeley National Laboratory,
Berkeley, CA~94720, USA;\\[-1mm]
\normalsize 
e-mail: {\tt JPGreensite@lbl.gov}\\
\normalsize 
$^{d}$Institute of Physics, Slovak Academy of Sciences,
SK--842 28 Bratislava, Slovakia;\\[-1mm]
\normalsize 
e-mail: {\tt fyziolej@savba.sk}}
\date{\small (November 5, 1999)}
\begin{document}
\maketitle
\thispagestyle{empty}
\begin{abstract}
The dominance of center degrees of freedom is observed in SU(3)
lattice gauge theory in maximal center gauge. The full asymptotic
string tension is reproduced, after center projection, by the
center elements alone. When center vortices are removed from
lattice configurations, the string tension tends to zero. 
This provides further evidence for the role played by center 
vortices in the mechanism of color confinement in quantum 
chromodynamics, but more extensive simulations with a better
gauge-fixing procedure are still needed. 
\end{abstract}
\vfill
\begin{flushright}
hep-lat/9911006
\end{flushright}
\section{Introduction}
The idea that center vortices play a decisive role in the
mechanism of color confinement in quantum chromodynamics was 
proposed more than 20 years ago by 't~Hooft~\cite{tHooft:1978}
and other authors~\cite{others}. 
Recently, the center-vortex picture of confinement has found remarkable
confirmation in numerical simulations of the SU(2) lattice gauge 
theory~\cite{DelDebbio:1997a,DelDebbio:1998a,DelDebbio:1998b,%
Faber:1999a,deForcrand:1999,Langfeld:1998,Engelhardt:1999,Bertle:1999,%
Bakker:1999,Kovacs:1999a}. 
Our group has proposed a technique for locating center vortices 
in thermalized lattice configurations based on fixing to the 
so called maximal center gauge, followed by center 
projection~\cite{DelDebbio:1997a,DelDebbio:1998a,DelDebbio:1998b}.  

In SU(2) lattice gauge theory, the maximal center gauge is a gauge 
in which the quantity
\beq
       R = \sum_x \sum_\m \Bigl| \mbox{Tr}[U_\m(x)] \Bigr|^2
\label{e1}
\eeq
reaches a maximum. This gauge condition forces each link variable to be
as close as possible, on average, to
a $Z_2$ center element, while preserving a residual $Z_2$ gauge invariance.
Center projection is a mapping of each SU(2) link variable to
the closest $Z_2$ center element:
\beq
        U_\m(x) \ra Z_\m(x) \equiv \mbox{signTr}[U_\m(x)].
\label{e2}
\eeq
The excitations on the projected $Z_2$ lattice are point-like, 
line-like, or surface-like objects, in $D=2,3$, or $4$ dimensions 
respectively, called ``P-vortices.''
These are thin objects, one lattice spacing across.
There is substantial numerical evidence that thin
P-vortices locate the middle of thick center vortices on the 
unprojected lattice. The string tension computed on center projected
configurations reproduces the entire asymptotic SU(2) string 
tension~\cite{DelDebbio:1998b}. 
It has also been demonstrated recently that removal of
center vortices not only removes the asymptotic string tension, but restores
chiral symmetry as well, and the SU(2) lattice is then brought to trivial 
topology~\cite{deForcrand:1999}.  The vortex density has been seen to scale 
as predicted by asymptotic 
freedom~\cite{Langfeld:1998,DelDebbio:1998b,Faber:1999a}. 
The properties of vortices have also been studied at finite 
temperature~\cite{Engelhardt:1999,Bertle:1999,Bakker:1999}, 
and it has been argued  
that the non-vanishing string tension of spatial Wilson loops 
in the deconfined phase can be understood in terms of vortices 
winding through the periodic time direction. 
We have also proposed a simple model which explains
the Casimir scaling of higher-representation string-tensions at 
intermediate distance scales in terms of the finite thickness of center 
vortices~\cite{Faber:1998}.

Putting the above and other pieces of evidence together, it seems clear
that our procedure of maximal-center-gauge fixing
and center projection identifies physical objects that play a crucial
role in the mechanism of color confinement. However, the gauge group
of QCD is color SU(3), not SU(2), and it is of utmost importance
to demonstrate that the observed phenomena are not specific to the SU(2)
gauge group only. Some preliminary results for SU(3) were presented
in Section 5 of Ref.\ \cite{DelDebbio:1998b}. They came from simulations on
very small lattices and at strong couplings. It was shown that 
center-projected Wilson loops reproduce results of the strong-coupling
expansion of the full theory up to $\beta\simeq 4$.

The purpose of the present letter is to present further
evidence on center dominance in SU(3) lattice gauge theory,
very similar to the results that arose from SU(2) simulations. Though
not as convincing as the SU(2) data, the first SU(3) results support the view
that the vortex mechanism works in SU(3) in the same way 
as in SU(2).

\section{Maximal Center Gauge in SU(3)}
The maximal center gauge in SU(3) gauge theory is defined as the gauge
which brings link variables $U$ as close as possible to elements of
its center $Z_3$. This can be achieved as in SU(2) by maximizing 
a ``mesonic'' quantity
\begin{equation}\label{mesonlike}
R=\sum_x\sum_\mu\mid\mbox{Tr}\;U_\mu(x)\mid^2,
\end{equation}
or, alternatively, a ``baryonic'' one
\begin{equation}\label{baryonlike}
R'=\sum_x\sum_\mu\mbox{Re}\left(\left[\mbox{Tr}\;U_\mu(x)\right]^3\right).
\end{equation}
The latter was the choice of Ref.\ \cite{DelDebbio:1998b}, 
where we used the method
of simulated annealing for iterative maximization procedure. The 
convergence to the maximum was rather slow and forced us to restrict
simulations to small lattices and strong couplings.

The results, that will be presented below, were obtained in a gauge 
defined by the ``mesonic'' condition (\ref{mesonlike}). The maximization
procedure for this case is inspired by the Cabibbo--Marinari--Okawa
SU(3) heat bath method~\cite{Cabibbo:1982,Okawa:1982}.%
\footnote{A similar approach was applied for SU(3) cooling 
by Hoek et al.~\cite{Hoek:1987}.}
The idea of the method is as follows: In the maximization
procedure we update link variables to locally maximize the quantity
(\ref{mesonlike}) with respect to a chosen link. At each site we thus 
need to find a gauge-transformation matrix $\Omega(x)$ which maximizes
a local quantity
\beq
R(x) =  \sum_{\mu}\left\{ 
\left\vert \mbox{Tr} \left[\Omega(x)^{\phantom{\dg}}
U_\mu(x) \right] \right\vert^2
+ \left\vert \mbox{Tr} \left[ 
U_\mu(x-\hat{\mu})\;\Omega^\dg(x) \right] \right\vert^2
\right\}.
\label{localcond}
\eeq
Instead of trying to find the optimal matrix $\Omega(x)$, we take an 
SU(2) matrix $g(x)$ and embed it into one of the three 
diagonal SU(2) subgroups of SU(3). The expression 
(\ref{localcond}) is then maximized with respect to $g$, 
with the constraint of $g$ being an SU(2) matrix.
This reduces to an algebraic problem (plus a solution of a non-linear
equation). Once we obtain the matrix $g(x)$, we update link variables
touching the site $x$, and repeat the procedure for all three subgroups
of SU(3) and for all lattice sites. This constitutes one center
gauge fixing sweep. We made up to 1200 sweeps for each configuration.
Center projection is then done by replacing the link matrix by the
closest element of $Z_3$.

The above iterative procedure was independently developed by
Montero, and described with full details in his recent publication
\cite{Montero:1999}. Montero, building on the work of 
Ref.~\cite{Gonzalez-Arroyo:1998}, has 
constructed classical SU(3) center vortex solutions on a periodic lattice.  
He has found that P-vortex plaquettes accurately locate the middle of the 
classical vortex, which is evidence of the ability
of maximal center gauge to properly find vortex locations. 

\vfill
\section{Center Dominance in SU(3) Lattice Gauge Theory}
The effect of creating a center vortex linked to a given Wilson loop
in SU(3) lattice gauge theory is to multiply the Wilson loop by an element
of the gauge group center, i.e.
\beq
W(C) \ra e^{\pm2\pi i/3} W(C).
\label{vortex_effect}
\eeq
Quantum fluctuations in the number of vortices linked to a Wilson loop
can be shown to lead to its area law falloff; the simplest, but 
urgent question is whether center disorder is sufficient to produce
the whole asymptotic string tension of full, unprojected lattice 
configurations.

\begin{figure}[b!]
\centerline{\includegraphics[width=10cm]{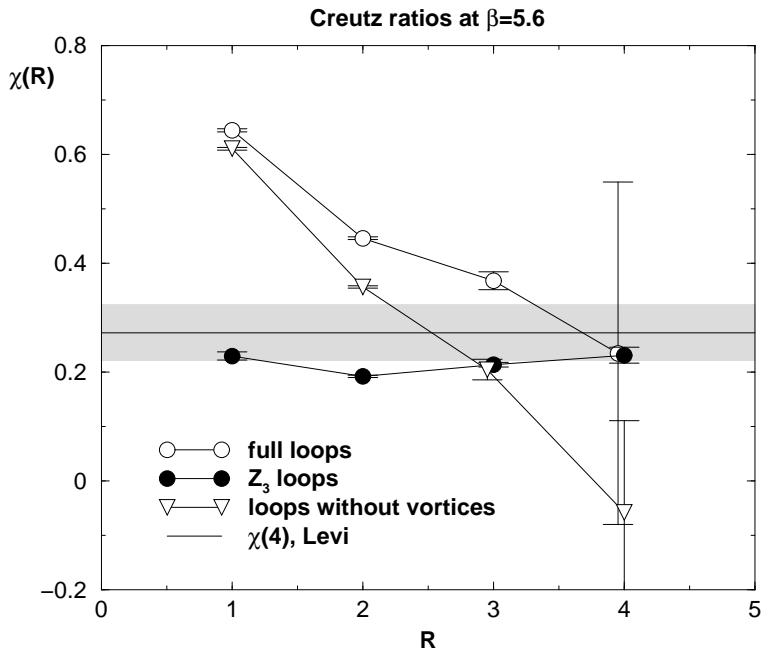}} 
\caption{Creutz ratios for the original, the $Z_3$ projected, 
and the modified (with vortices removed) ensembles. 
($\beta=5.6$, $12^4$ lattice.) For comparison, the value of
$\chi(4)$ and its error is shown in grey. 
The value comes from the compilation of Levi~\protect\cite{Levi}.}
\label{chi_vs_R}
\end{figure}

We have computed Wilson loops and Creutz ratios at various values
of the coupling $\beta$ on a $12^4$ lattice, from full lattice
configurations, center-projected link configurations in
maximal center gauge, and also
from configurations with all vortices removed. Figure \ref{chi_vs_R}
shows a typical plot at $\beta=5.6$. It is obvious that center
elements themselves produce a value of the string tension which
is close to the asymptotic value of the full theory. On 
the other hand, if center elements are factored out from link matrices
and Wilson loops are computed from SU(3)/$Z_3$ elements only, 
the Creutz ratios tend to zero for sufficiently large loops.
The errorbars are, however, rather large, and one cannot
draw an unambiguous conclusion from the data.

Recently we have argued that center dominance by itself does
not prove the role of center degrees of freedom in QCD 
dynamics~\cite{Ambjorn:1998,Faber:1999a};
some sort of center dominance exists also without any gauge fixing
and can hardly by attributed to center vortices. Distinctive
features of center-projected configurations in
maximal center gauge in SU(2), besides center dominance, were that:\\ 
1.\ Creutz ratios were approximately constant starting from
small distances (this we called ``precocious linearity''),\\
2.\ the vortex density scaled with $\beta$ exactly as expected
for a physical quantity with dimensions of inverse area.
 
Precocious linearity, the absence of the Coulomb part of the
potential on the center-projected lattice at short distances,
can be quite clearly seen from Fig.\ \ref{chi_vs_R}. One observes
some decrease of the Creutz ratios at intermediate distances. A similar effect
is present also at other values of $\beta$. It is not clear to us
whether this decrease is of any physical relevance, or whether it should be
attributed to imperfect fixing to the maximal center gauge.

\begin{figure}[b!]
\centerline{\includegraphics[width=10cm]{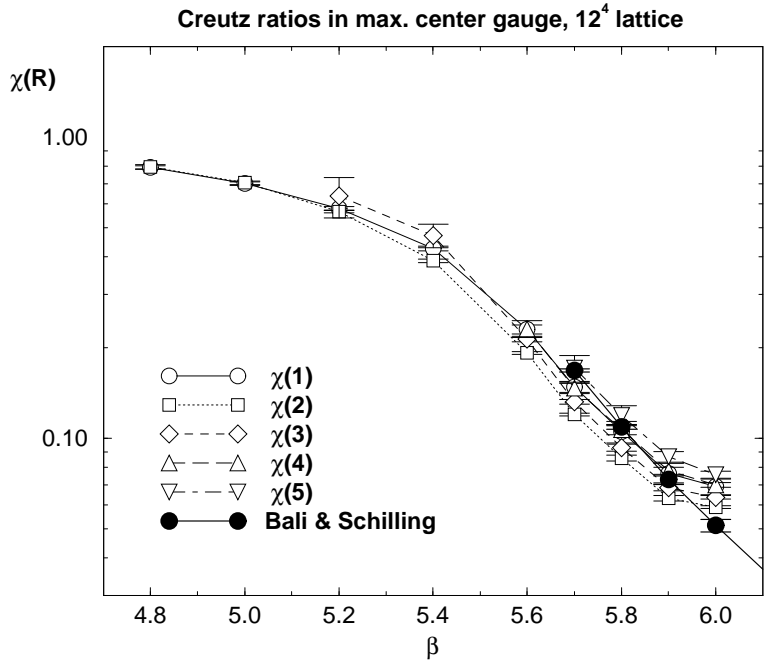}} 
\caption{Center-projected Creutz ratios vs.\ $\beta$. Full
circles connected with a solid line are asymptotic values 
quoted by Bali and Schilling~\protect\cite{Bali:1993}.}
\label{chi_vs_b_with_bali}
\end{figure}

The issue of scaling is addressed in Figure \ref{chi_vs_b_with_bali}.
Here values of various Creutz ratios are shown as a function of $\beta$
and compared to those quoted in Ref.\ \cite{Bali:1993}. All values
for a given $\beta$ lie close to each other (precocious linearity 
once again) and are in reasonable agreement with asymptotic values obtained
in time-consuming SU(3) pure gauge theory simulations. The plot in 
Fig.\ \ref{chi_vs_b_with_bali} is at the same time a hint
that the P-vortex density also scales properly. 
The density is approximately proportional 
to the value of $\chi(1)$ in center-projected configurations, and 
$\chi(1)$ follows the same scaling curve as Creutz ratios obtained
from larger Wilson loops. 

A closer look at Fig.\  \ref{chi_vs_b_with_bali} reveals that
there is no perfect scaling, similar to the SU(2) case, in our SU(3)
data. Broken lines connecting the data points tend to bend
at higher values of $\beta$. In our opinion, this is a finite-volume
effect and should disappear for larger lattices. In our simulations
we use the QCDF90 package~\cite{QCDF90} (supplemented by subroutines for
MCG fixing and center projection), which becomes rather inefficient
on a larger lattice~\cite{Rebbi:1999}. 
CPU and memory limitations do not allow us 
at present to extend simulations to larger lattice volumes.

An important test of the vortex-condensation picture is the
measurement of vortex-limited Wilson loops. Let us denote 
$\langle W_n(C)\rangle$ the expectation value of the Wilson loop
evaluated on a sub-ensemble of {\em unprojected\/} lattice configurations,
selected such that precisely $n$ P-vortices, in the corresponding
{\em center-projected\/} configurations, pierce the minimal area of the loop.
For large loop areas one expects
\beq
\frac{\langle W_n(C)\rangle}{\langle W_0(C)\rangle}\ra e^{2n\pi i/3}.
\label{Wn/W0}
\eeq
We tried to measure quantities like 
${\langle W_1(C)\rangle}/{\langle W_0(C)\rangle}$ and 
${\langle W_2(C)\rangle}/{\langle W_0(C)\rangle}$ 
in Monte Carlo simulations. The trend of our data is in 
accordance with the expectation based on the $Z_3$ vortex-condensation
theory, Eq.\ (\ref{Wn/W0}), but before the evidence becomes conclusive,
the errorbars become too large.   

\section{Conclusion}
We have presented evidence for center dominance, precocious linearity,
and scaling of center-projected Creutz ratios and of P-vortex density
from simulations of the SU(3) lattice gauge theory. Our data -- and conclusions
that can be drawn from them -- look quite similar to the case of SU(2).
However, the SU(3) data at present are not as convincing and unambiguous
as those of SU(2), the errorbars are still quite large and much more
CPU time would be required to reduce them. The reason essentially is
that the gauge-fixing maximization for SU(3) is very time consuming,
either with simulated annealing or by the Cabibbo--Marinari--Okawa-like
method used in the present investigation.%
\footnote{Typically thousands of iterations were needed for gauge
fixing also in the investigation of Montero \cite{Montero:1999}.}
Moreover, the maximal center gauge is known to suffer from the
Gribov problem, which makes gauge fixing notoriously difficult
(in this context see also Refs.~\cite{Kovacs:1999b,Faber:1999b}).
A better alternative is badly needed, and may be provided 
by the recent proposal of de Forcrand et al.~\cite{Alexandrou:1999} 
based on fixing to the so-called Laplacian center gauge. 
Their first SU(2) results are promising, and the method can readily 
be extended to the case of SU(3).

It is encouraging that none of the pieces of data, which
we have accumulated in SU(3) lattice gauge theory until now,
contradicts conclusions drawn from earlier SU(2) results.
If future extensive simulations with a more suitable, Gribov-copy
free center-gauge fixing method confirm the evidence obtained
in our exploratory investigation, center vortices  
will have a very strong claim to be the true mechanism of color confinement
in QCD.  

\subsection*{Acknowledgements}   
{Our research is supported in part by Fonds zur F\"orderung der
Wissenschaftlichen Forschung P13997-PHY (M.F.), the U.S. Department of 
Energy under Grant No.\ DE-FG03-92ER40711 (J.G.), and the Slovak Grant 
Agency for Science, Grant No. 2/4111/97 (\v{S}.O.). In earlier stages
of this work \v{S}.O.\ was also supported by the ``Action 
Austria--Slovak Republic: Cooperation in Science and Education''
(Project No.\ 18s41). Portions of our numerical simulations were 
carried out on computers of the Technical University of 
Vienna, and of the Computing Center of the Slovak Academy of Sciences 
in Bratislava.}

\end{document}